\documentclass[12pt]{article}
\usepackage{graphicx}
\usepackage{axodraw}
\usepackage{epic}
\usepackage{amssymb,amsmath}
\def\beq#1\eeq{\begin{equation}#1\end{equation}}    
\def\bea{\begin{eqnarray}}  
\def\eea{\end{eqnarray}}  
\def\bq{\begin{quote}}  
\def\eq{\end{quote}}  
\def\bi{\begin{itemize}}  
\def\ei{\end{itemize}}  
\def\be{\begin{enumerate}}  
\def\ee{\end{enumerate}}  
\def\bi{\begin{itemize}}
\def\ei{\end{itemize}} 
\def\bc{\begin{center}}
\def\ec{\end{center}}  
\def\pa{\partial}

\def\de{\delta}
  
\def\cl{{\cal L}} 
 
\def\La{\Lambda}

\def\r2{\sqrt{2}} \def\rt{\sqrt{2}} 
\def\ra{\rightarrow}
\def\bi{\begin{itemize}}  
\def\ei{\end{itemize}}    
\def\dy{ \Delta y}

\def\nn{\nonumber \\}  
\def\t{\tilde}

 \def\no{{\cal N}=1}
\def\nt{{\cal N}=2}
\def\im{\item}

\def\d{{\rm d}}
\def\Tr{{\rm Tr}}
\def\ZZ{\mathbb{Z}}

\newsavebox{\moose}
\sbox{\moose}{%
\begin{picture}(0,0)
  \thicklines
  \put(-60,0){\circle{35}}
  \put(60,0){\circle{35}}
  \ArrowLine(-40,0)(40,0)
\end{picture}
}

\newsavebox{\site}
\sbox{\site}{%
\begin{picture}(0,0)
  \thicklines
  \put(60,0){\circle{35}}
  \ArrowLine(-40,0)(40,0)
  \ArrowLine( 80,-60)(60,-20)
  \ArrowLine( 60,-20)( 40,-60)
\end{picture}
}

\newcommand{\NPB}[3]{\emph{ Nucl.~Phys.} \textbf{B#1} (#2) #3}   
\newcommand{\PLB}[3]{\emph{ Phys.~Lett.} \textbf{B#1} (#2) #3}   
\newcommand{\PRD}[3]{\emph{ Phys.~Rev.} \textbf{D#1} (#2) #3}

\newcommand{\JHEP}[3]{\emph{JHEP} \textbf{#1} (#2) #3}

\begin{document}
\pagestyle{empty}
\setcounter{page}{0}
{\normalsize\sf
\rightline {hep-ph/0212206}
\rightline {IFT-02/42}
\vskip 3mm
\rm\rightline{December 2002}
}

\vskip 1.0cm
\begin{center}
{\huge \bf 
Deconstructing 5D supersymmetric \\ 
$U(1)$ gauge theories on orbifolds
}\\
{\huge \bf }
\vspace*{1cm}
\end{center}
 \noindent
\vskip 0.5cm
\centerline
{\sc Adam  Falkowski ${}^{1}$,
 Hans-Peter Nilles  ${}^{2}$,}
\centerline{\sc Marek Olechowski ${}^{1}$, 
Stefan  Pokorski ${}^{1}$}
\vskip 1cm
\centerline{\em ${}^{1}$ Institute of Theoretical Physics, 
Warsaw University}
\centerline{\em Ho\.za 69, 00-681 Warsaw, Poland}
\vskip.2cm
\centerline {\em  ${}^{2}$ Physikalisches Institut der 
Universit\"at Bonn}
\centerline {\em Nussallee 12, 53115 Bonn, Germany}
\vskip.3cm
\centerline{\tt \small  afalkows@fuw.edu.pl, 
nilles@th.physik.uni-bonn.de,}
\centerline{\tt \small olech@fuw.edu.pl, pokorski@fuw.edu.pl}
\vskip 1.5cm

\centerline{\bf Abstract}
\noindent
We investigate deconstruction of five dimensional supersymmetric abelian gauge theories compactified on $S_1/\ZZ_2$, with various sets of bulk and matter multiplets. The problem of anomalies, chirality and stability in the deconstructed theories is discussed. We find that for most of the 5d brane/bulk matter assignments there exists the deconstructed version. There are, however, some exceptions. 

\vskip .3cm

\newpage

\setcounter{page}{1} \pagestyle{plain}


Higher dimensional gauge theories offer interesting new tools to
understand the roots of the Standard Model. Among other things, 
compactification on orbifolds is a very efficient mechanism of
reducing symmetries. Moreover compactification on orbifolds is a
simple mechanism to generate chirality in four dimensions. Another
important virtue of higher dimensional theories is the possibility of
localizing wave functions in extra dimensions. This can explain the
hierarchy of various physical parameters, e.g. fermion masses,  as a
result of a small overlap of wave functions localized at different
positions in extra dimensions. However, gauge theories in more than
four dimensions are non-renormalizable and some quantum problems
cannot be addressed in an unambiguous way.

It has recently been demonstrated 
\cite{ARCOGE_dec,HIPOWA}
that the physics of higher
dimensional gauge theories can be reproduced in certain four
dimensional theories with enlarged gauge symmetry. For example,
the correspondence exists between  five dimensional gauge theories
with the gauge group $G$ and  four dimensional gauge theories with the
gauge group $G$ replicated $N$ times,  $G \times G \times \dots \times
G$. The four dimensional theory is referred to as `latticized' or
`deconstructed' and can be viewed as a renormalizable completion of
the latter. A more general view on deconstruction is that, inspired by
higher dimensional gauge theories, one arrives at a class of purely 4d
renormalizable gauge theories that offer (and often generalize)
similar benefits to those of higher dimensional gauge theories.

Recently, some attention has been focused on 5d supersymmetric $U(1)$
gauge theories compactified on the orbifold $S_1/\ZZ_2$ (or more
generally on the question of anomalies, localization and stability
which is most easily studied in the $U(1)$ case). In this letter we
investigate how similar properties appear in deconstruction.  As a
by-product of this discussion we clarify some aspects of the
correspondence between the geometrical space in 5d  and the `group
product space' in 4d.

We begin with recalling the construction of  5d supersymmetric $U(1)$
gauge theories on orbifolds. The 5d Abelian gauge theory in a {\it
flat} background in the ${\cal N}=1$ superspace formalism
\cite{MASASI,ARGRWA,MAPO,HE} reads: 
\beq
\label{eq.sfga}
S_{5g} = \int \d^4x \d y 
\left\{\left [ \frac{1}{2} W^\alpha W_\alpha + {\rm h.c.}\right ]_F 
+\left[\left(\pa_5 V - \frac{1}{\rt}\left(\Phi + \Phi^\dagger\right)
\right)^2 \right]_D 
\right\}\,.
\eeq  
In the above, $V = (A_\mu,\chi,D)$ is the $\no$ vector multiplet and 
$\Phi = (\frac{1}{\rt}\Sigma + i\frac{1}{\rt} A_5,\Psi,G)$ is a chiral
multiplet, singlet under $U(1)$, which completes the vector multiplet
to the 5d $\nt$ multiplet.

The action for a  bulk matter multiplet (called hypermultiplet)
charged under $U(1)$ is given by: 
\beq
\label{eq.fha}
S_{5h} = \int \d^4x \d y \left [ H^\dagger e^{2 g_5 q V} H 
+ \tilde H^\dagger e^{-2 g_5 q V}\tilde  H \right ]_D  
+ \int \d^4 x\d y \left [\sqrt{2} g_5 \tilde H \Phi H  
+  \tilde H \pa_5 H + {\rm h.c.}\right]_F
\eeq
where $H = (H,\psi,F)$ and $\tilde H =(\t H, \t\psi,\t F) $ are two
chiral multiplets in fundamental and anti-fundamental representation
of the gauge group that make up one 5d hypermultiplet.

The pure 5d supersymmetric $U(1)$ gauge theory on $S_1/\ZZ_2$ is
non-anomalous. This is because all fields of the gauge multiplet are
$U(1)$ singlets.
But in models with a charged hypermultiplet  the orbifold projection
leaves only one chiral zero-mode and so  the 4d effective theory is
anomalous. This anomaly manifests itself in a peculiar way in the full
5d set-up, namely,  {\it half} of the anomaly is localized at each
fixed point \cite{ARCOGE_ano}: 
\beq
\label{e.bha}
\pa_\alpha J ^\alpha  
= 
\frac12 \left[\delta (y) + \delta (y -\pi R)\right]{\cal Q} 
\eeq
where ${\cal Q}$ is the standard anomaly of the 4d effective theory
(analogous anomalies for more complicated orbifolds are discussed in
ref.\ \cite{GGNNOW}).
This anomaly can of course be cancelled by adding another
hypermultiplet with the zero mode of opposite charge. Another option
is to add a chiral multiplet of opposite charge at one of the fixed
points which also contribute to localized anomalies. Both
possibilities lead to non-anomalous zero-mode spectrum  but in the
latter case, due to the factor $1/2$ in eq.\ (\ref{e.bha}), the 5d
current still looks anomalous,    
$\pa_\alpha J ^\alpha=\frac12 [-\delta(y)+\delta(y -\pi R)]{\cal Q}$.
However, this would-be anomaly can be removed by adding
a local Chern-Simons counterterm \cite{SSSZ,PIRI,BCCRS} 
and does not lead to any
inconsistencies of the theory. Hence, to have a non-anomalous 5d model
of this type it is enough to insist on non-anomalous spectrum of the
zero-modes.

It is well-known that 4d supersymmetric theories with $U(1)$ gauge
symmetry allow for the presence of the $\xi[V]_D$ term in the
action. In 5d models the situation is different as the symmetries
($\nt$ in the bulk and $\no$ on the boundaries) allow only for FI
terms localized at the boundaries \cite{GGNN}, 
$\int\d^4x\d y D[\xi_0\delta (y) +\xi_\pi\delta (y - \pi R)]$.
If we insist that supersymmetry is not spontaneously broken, the
vacuum configuration must satisfy the D-flatness condition:  
\beq
\int\d y
\left[\xi_0\delta(y) + \xi_\pi\delta(y-\pi R)
+ g_5q(H^2  - \t H^2)\right]  
=0 
\,.
\eeq
If the gauge symmetry is to stay unbroken, the hypermultiplet scalars
cannot receive any vevs. Then, from the D-flatness condition it
follows that the FI terms satisfy:   
\beq
\label{e.fic} 
\xi_0 + \xi_\pi = 0
\,.
\eeq
The condition (\ref{e.fic}) translates into vanishing of an FI term
in the 4d effective theory. The effect of such FI term in the 5d
picture  is to induce an expectation value of the gauge multiplet
scalar $\Sigma$ according to the equation: 
\beq  
\label{e.kinkmass5d}
\langle\Sigma\rangle = \frac12 \xi_\pi \epsilon(y) 
\,.
\eeq
If only the gauge multiplet were present, the vev of $\Sigma$ would
have no effect whatsoever on the low-energy effective theory. In the
presence of a bulk hypermultiplet, the vev of $\Sigma$ induces  the
hypermultiplet {\it kink-mass} term: 
\beq
W = \rt g_5 \Phi H \tilde H \ra  M \epsilon(y) H \tilde H
\eeq
with $M = \frac12 g_5\xi_\pi $. Such a kink-mass leaves the zero-mode
massless  while it shifts the tower of the massive KK modes, $m_n^2 =
M^2 + (n/R)^2$ for $n>0$. It also  disturbs the profiles of the wave
functions, in particular, it leads to an exponential localization of
the zero-mode: 
\beq
H^{(0)} = \sqrt{\frac{M}{2\left(e^{ M \pi R} -1\right)}}\, e^{M |y|}  
\,. 
\eeq
Depending on the sign of $M$, the zero-mode is  localized either on
the $y=0$ or $y= \pi R $ brane.

It was found in ref.\ \cite{SSSZ,GRNIOL} that in 5d localized FI terms can
be generated dynamically. More precisely a bulk hypermultiplet with
the zero-mode of charge $q$ generates the operator  
\begin{equation}
D g\frac{q}{2}
\left\{
\frac{\La^2}{16\pi^2}
\left[\de(y)+\de(y-\pi R)\right]
+\frac{\ln\La^2}{64\pi^2}\left[\de''(y)+\de''(y-\pi R)\right]
+\ldots
\right\}.
\label{e.fib}
\end{equation}
{It has contributions localized at the orbifold fixed points which are 
quadratically  sensitive to the cut off scale $\La$. Besides, there are logarithmically divergent contributions that depend on the thickness $\sigma$ of the brane. 
If $\sigma$ is of the order of  $\Lambda^{-1}$ then we expect that the $\de''$ terms are subleading.
}
On the other hand a brane chiral multiplet of charge $q_0$,
located at $y=0$, generates just the standard FI term 
$Dgq_0\frac{\La^2}{16\pi^2}\de(y)$ localized at $y=0$. 
This raises the question about stability of various configurations of
matter fields in the 5d set-up. As we discussed previously we should
concentrate on those configurations for which the sum of charges ${\rm Tr} \, q$ of the
massless modes vanishes. Thus we can consider the following examples
of just two massless fields with opposite charges: 
\bi 
\im 
{\it Two bulk hypermultiplets with the zero-modes of opposite
charges}. This configuration is perfectly stable, as the  operators of
eq.\ \ref{e.fib} generated by the two hypermultiplets cancel.    
\im 
{\it Two 4d chiral multiplets of opposite charges localized at the
fixed points}. If both multiplets live at the same fixed point, the
generated FI terms of course cancel. If they live at different fixed
points then localized FI terms satisfying the condition (\ref{e.fic})
are generated.
\im 
{\it One bulk hypermultiplet of charge $q$ together with one brane chiral
multiplet of charge $-q$ localized at $y=0$}. In such a case the
operator 
\beq
D g\frac{q}{2}
\left\{
\frac{\La^2}{16\pi^2}
\left[-\de(y)+\de(y-\pi R)\right]
+\frac{\ln\La^2}{64\pi^2}\left[\de''(y)+\de''(y-\pi R)\right]
\right\}
\label{eq:FI}
\eeq
is generated. As a result, the profile of the hypermultiplet zero-mode
is modified so that for large $\La$ it is sharply localized at $y=0$
where the chiral multiplet lives. This spontaneous localization is
partly due to the kink-mass for the hypermultiplet generated by
the terms in (\ref{eq:FI}) proportional to $\La^2$ (such terms alone
would lead to an exponential profile). 
On top of it, the $\de''$ terms may lead to further localization of the
zero-mode in a region around $y=0$ with the thickness $\sigma$ given by that of
the fixed point brane.

In addition to the localization of the zero mode, the massive 
Kaluza--Klein modes become very heavy with masses above the cut off
scale. In such a way a bulk field effectively becomes a brane
field: it is localized at a brane and it has no massive modes
(below the cut off).
\ei

The above simple example can be generalized to more complicated
situations with more than two multiplets. 
In general there can be three types of fields:
brane chiral multiplets at $y=0$ with the $U(1)$ charges $q_0$;
brane chiral multiplets at $y=\pi R$ with charges $q_\pi$;
bulk hypermultiplets with zero mode charges $q_B$. Such a model is
anomaly free and can have unbroken supersymmetry if
\beq
\sum q_0 + \sum q_\pi + \sum q_B = 0
\,.
\label{eq:anomaly5}
\eeq
Not all such models are stable. In some cases the bulk fields get
localized and effectively change to brane fields. { It has been shown in
ref.\ \cite{GRNIOL} that the bulk fields are stable if the zero-mode charges sum
up to zero not only globally but also locally:
\beq
\sum q_0 + \frac12\sum q_B = 0\,, 
\qquad\qquad
\sum q_\pi + \frac12\sum q_B = 0\,.
\label{eq:stability5}
\eeq
These  two conditions correspond to the part of the FI
terms (\ref{e.fib}) proportional to $\Lambda^2$. In case the $\delta''$ terms are not subleading  we must ensure cancellation of logarithmically divergent contributions to the FI terms. Then there is an additional stability condition:
\beq
\sum q_B = 0\,.
\label{eq:stability5s}
\eeq}

Later we will compare 5d models with models obtained in
deconstruction. Such comparison should be performed at the level of
the effective 4d models. Thus, one needs a criterion to distinguish
the bulk fields from the brane fields from the 4d point of view. The
existence of massive KK modes is such a criterion: a brane field has
only the zero mode while a bulk field has a zero mode and a tower of
massive modes (with masses below the cut-off or the deconstruction scale).

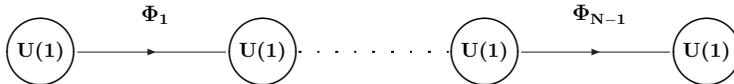
\begin{figure}[tb] 
\centering
  \scalebox{.70}{%
\begin{picture}(-100,100)(0,0)
      \thicklines
      \put(-180,50){\usebox\moose}
      \put(60,50){{\usebox\moose}}
      \put(-60,50){{\makebox(0,0){\dottedline{10}(-40,0)(40,0)}}}
      \put(-240,50){\makebox(0,0){\small\bf U(1)}}
      \put(-120,50){\makebox(0,0){\small\bf U(1)}}
      \put(0   ,50)   {\makebox(0,0){\small\bf U(1)}}
      \put( 120,50){\makebox(0,0){\small\bf U(1)}}
      \put(-180,70){\makebox(0,0){ $\bf \Phi_1$}}
      \put(  60,70){\makebox(0,0){ $ \bf \Phi_{N-1}$}}
    \end{picture}
} 
\caption{The quiver diagram of the model}
\label{f.qd}
\end{figure}

In the remainder of this  letter we discuss the issue of FI terms,
anomalies and localization in the deconstruction set-up.  It has been
suggested in ref.\ \cite{HIPOWA} (and in \cite{CSERGR} for a
supersymmetric case) that the physics of gauge theories on orbifolds
can  be realized in deconstruction if the quiver diagram of the 4d
model is of the `aliphatic' type,  see Fig.\ \ref{f.qd}. More precisely,
deconstruction of 5d supersymmetric $U(1)$ gauge theory involves $N$
$U(1)$ gauge multiplets $V_p$ and $N-1$ chiral multiplets $\Phi_p$
(called link-Higgs) charged as $(Q,-Q)$ under the $p$-th and $p+1$-th
gauge group, respectively. Note that such choice of the  charges
introduces `orientation' in the group product space. The vacuum
expectation values of the link-Higgs bosons break the product group
down to the diagonal subgroup and it is below the scale set by these
vevs where the correspondence holds.

The correspondence to gauge theories in a flat background is realized
by  assuming universal values of the  gauge coupling and link-Higgs
vevs, $g_p=g$, $v_p = v$ (nonuniversal values correspond to 5d gauge
theories in warped backgrounds \cite{ABKOMA,FAKI,RASHWE}).
For deconstruction of  $SU(M)$ gauge
theories, arbitrary link-Higgs vevs are flat directions of the scalar
potential. This is no longer the case for deconstructing $U(1)$.  
Note first that now  FI terms for every gauge group are  consistent
with the symmetries, as in deconstruction we have only $\no$
supersymmetry. Adding the FI terms $ \sum_p 2 \xi_p [V_p]_D$ results
in the scalar potential:  
\beq
V = \frac12 g^2 \left [
(Q|\Phi_1|^2 + \xi_1)^2 + (Q|\Phi_2|^2  - Q|\Phi_1|^2 + \xi_2)^2 + \dots
+ (- Q|\Phi_{N-1}|^2 + \xi_{N})^2 \right ].
\eeq      
The first thing to see here is that if  all the FI terms were set to
zero, the minimum of this potential would be   at $\langle \Phi_p
\rangle =0$ (corresponding to an unbroken product gauge group)  and
there would be no energy range where the deconstruction model could
match the 5d gauge theory. This situation is different from the 5d
case, where the presence of FI terms is by no means
necessary. Secondly, the minimum with unbroken supersymmetry
satisfies:  
\bea
Q\langle \Phi_1 \rangle^2  &=& -{\xi_1}\,,\nn
Q\langle\Phi_2 \rangle^2 &=& -{\xi_1-\xi_2}\,,\nn
&\dots& \nn 
Q\langle\Phi_{N-1}\rangle^2 &=& -{\xi_1-\xi_2-\ldots-\xi_{N-1}} 
= {\xi_N}\,.
\eea
As we are interested here in models with universal link-Higgs vevs we
must further constrain $\xi_2 = \ldots 
=\xi_{N-1} =0$, thus we must forbid the appearance of FI terms in all
except the boundary gauge groups. In such a case the existence of a
supersymmetric minimum requires the  
(fine-tuning) condition on  the FI terms and additional conditions on
their signs:
\bea
\xi_1 + \xi_N &=& 0
\,,
\nn
Q \xi_1 &<& 0
\,.
\label{eq.smc}
\eea
(From the above it follows that $Q \xi_N > 0$.)
The former condition is clearly the analog of eq.\ (\ref{e.fic}) which 
ensures that the FI term in the effective low-energy theory vanishes. 
The latter has no corresponding condition  in the 5d theory, which
again signals that the role of FI terms in deconstruction cannot be
exactly mapped on the 5d theory. At this stage the FI terms are
introduced `by hand'. Once their magnitude is  chosen, the
deconstruction scale is unambiguously determined, thus the
arbitrariness in choosing the FI terms translates into arbitrariness
of choosing the cut-off scale of the 5d theory.    
Observe that the FI terms generated by the link fields (which give
nonzero net charges for the first and the last groups) can not be used
to break the product gauge group because they do not satisfy the
second of the conditions (\ref{eq.smc}).

 5d gauge theories on $S_1/Z_2$, without matter, are of course anomaly free as all the fields in the gauge supermultiplet are $U(1)$ singlets.
However  in deconstruction the link-Higgs multiplets are chiral and are charged under $U(1)$, which implies that the problem of  anomalies has to be reconsidered. 
Indeed, the  deconstruction model as it stands is  inconsistent as the $U(1)$ gauge symmetries are anomalous. There are two kinds of anomalies:  mixed anomalies \cite{dufapo} of the neighboring groups and  boundary anomalies \cite{CSERGR}, that is the anomalies of the first and the $N$-th group. The anomalous variation of the action can be written in the superspace formalism  as:
 \bea&
\label{e.mixedanomalyaliphatic}
\delta \cl_{an} 
=  - \frac{i}{12 \pi^2} Q^2 \sum_{p=1}^{N-1} \int d^2 \theta \
\Lambda_p \
\left  (  W_{p+1}^{\alpha} W_{\alpha, p+1}
-  W_{p-1}^{\alpha} W_{\alpha,p-1} 
 - 2 W_{p}^{\alpha} W_{\alpha, p+1} 
 + 2 W_{p}^{\alpha} W_{\alpha,p-1} \right )
& \nn & 
- \frac{i}{12 \pi^2} Q^2 \int d^2 \theta \  \left ( \Lambda_1 \
      W_{1}^{\alpha} W_{\alpha,1}
 - \Lambda_N \  W_{N}^{\alpha} W_{\alpha,N} \right)  + {\rm h.c} \ , 
\eea
This variation can be cancelled by adding a local polynomial in the link-Higgs and the gauge fields, the so-called Wess-Zumino (WZ) terms. In  deconstruction we should impose an additional constraint that, in the continuum limit,  the WZ terms match some 5d invariant term. A natural candidate \cite{SKSM} for the continuum limit is the 5d Chern-Simons (CS) term.  With such constraint,   
in order to cancel the mixed anomalies we can choose 
any of the family of WZ terms parametrized by $C$: 
\bea
\label{e.superwz}
&
\cl_{SWZ} = -\frac{1}{24 \pi^2} Q^2 \int d^2\theta \log (\Phi_p/v) \left [
 \right . & \nn &  \left .
 (C-1) W_{\alpha,p}W^{\alpha}_p + (C-1)  W_{\alpha,p+1}W^{\alpha}_{p+1} 
+ (C+2) W_{\alpha,p}W^{\alpha}_{p+1}  \right ]
& \nn & 
- \frac{C}{24 \pi^2} Q^2 \int d^4\theta\left [ 
 (V_p D_\alpha V_{p+1} - V_{p+1} D_\alpha V_{p})(W^{\alpha}_p +W^{\alpha}_{p+1})  
\right ]
+{\rm h.c.}& \,
\eea 
that in the continuum limit yield the 5d CS term 
$\cl_{CS} = -\frac{C Q^2}{12 \pi^2}
\epsilon _{\alpha\beta\gamma\delta\epsilon} \left [
 A_\alpha \pa_\beta A_\gamma \pa_\delta A_\epsilon \right] $. However cancellation of the boundary anomalies uniquely sets $C=0$ in eq. (\ref{e.superwz}). Therefore the continuum limit of the deconstructed theory is a 5d theory with no CS term, as it should be in the absence of matter. 

 Note however, that cancelling anomalies via the WZ terms leaves
non-zero $\Tr q$ in the boundary groups, which generates additional  FI terms at the one-loop level. The resulting shift of  the FI terms may result in  instability  of the model. 
We shall return to the question of stability of various deconstructed 5d configurations at the end of this letter.        

Now we turn our attention to the physics of a 5d hypermultiplet
realized in deconstruction. In order to mimic hypermultiplets  one
introduces \cite{CSERGR} two sets of  chiral multiplets, $H_p=
(H_p,\psi_p)$ and $\tilde H_p =(\tilde H_p,\tilde \psi_p)$
(later called `replicated multiplets'), 
with charge $Q$ and $-Q$ with respect to the p-th gauge group. 
The most general renormalizable superpotential with mass terms and
couplings independent of $p$ is the following  
\beq
\label{eq.gs}
W = \sum_{p=1}^{N-1}\sqrt{2} \lambda \tilde H_p \Phi_p H_{p+1}
-\sum_{p=1}^{N} m \tilde H_p H_p \,.  
\eeq

The Yukawa coupling $\lambda$ should be fine-tuned to the gauge
coupling, $\lambda =g$, in order to match the $\nt$ supersymmetric
interactions of the 5d theory. Furthermore, the fine-tuning of the
mass parameter $m$ to the link-Higgs vev, $m = g v $, leads to the
similar spectrum and interactions as those of a 5d bulk hypermultiplet
without a kink-mass term. Note also that, with such  set of
links, we can only reproduce hypermultiplets with charge $Q$. If we allow for non-renormalizable interactions in the superpotential all rational charges are allowed.

When deconstructing orbifold theories, one has to set either 
$\tilde H_N \equiv 0$ or $H_1 \equiv 0$. The first choice results in
the zero mode of charge  $Q$ under the diagonal group while the
second yields the zero mode of charge $-Q$. This way one introduces
chirality in the matter sector.  
In 5d gauge theories chirality appears due to the $\ZZ_2$ symmetry 
(or boundary conditions in the `downstairs picture') which removes
some of the degrees  
of freedom from the spectrum. $\ZZ_2$ acts differently on left- and
right-handed fermion 
 components and, in particular, it leaves in the spectrum only one
chiral component of the zero mode. In deconstruction we have neither
$\ZZ_2$ nor boundaries to define the boundary  
conditions and chirality must be introduced `by hand'.  
For the case without matter multiplets this step is fairly
straightforward. 
Going from the periodic to the aliphatic quiver diagram consists in
removing one chiral multiplet $\Phi_N$  which results in  an odd
number of chiral fermions in the theory ($N$ gauginos and $N-1$ link
Higgsinos). Thus the intuitive  
step of turning the `circle' into a `line' automatically introduces
chirality as well. 
When the replicated matter fields $H_p$, $\tilde H_p$
are present one can  apparently remove many different chiral
multiplets to  
introduce chirality. However these possibilities  are not equivalent
and {\it only}  
removing $\tilde H_N$ or $H_1$ yields, in the deconstruction phase,
the spectrum and interactions similar to those of the 5d case. For
example, the other intuitive possibilities -  removing $\tilde H_1$
or $H_N$ -  yield the low-energy spectrum which does not correspond to
any 5d model. The difference can be seen already at the level of the
superpotential (\ref{eq.gs}). By removing $\tilde H_N$ (or $H_1$) we
remove just one mass term from the superpotential while by removing 
$H_N$ (or $\tilde H_1$) we remove one mass term and also one Yukawa
coupling term.

Thus the two `boundaries' of the group product space are by no means
equivalent. This is obviously counterintuitive to any geometric
interpretation of the group product space. Moreover it once again
illustrates the fact that, in case of orbifold theories, the
correspondence between 5d and deconstruction holds at the level of the
effective low-energy theory only for some specific choices made during
construction of the model.

Removing the anti-fundamental chiral multiplet $\tilde H_N$ leads to 
an anomaly localized at the $N$-th site of the group product space,
similarly, removing $H_1$ yields an anomaly of $U(1)_1$. Below the
deconstruction scale the anomaly of the diagonal group {\it globally}
matches the anomaly of the effective model obtained from the 5d theory
with one hypermultiplet. But above 
that scale the situation looks different than in  5d, where half of
the  anomaly is localized at each fixed point. This difference 
originates from the fact that in the deconstruction set-up 
the group product space is oriented and gives an oriented line in the
aliphatic case while both fixed points of the orbifold are
equivalent.

It is possible to  calculate the spectrum and the mode decomposition
of the matter  fields  $H$ and $\tilde H$.  
The case with $m=gv$ was studied in \cite{CSERGR} and the
correspondence to the 5d {\it massless} bulk hypermultiplet spectrum
was shown. We shall see that  $m \neq gv$ corresponds to a bulk
hypermultiplet with a kink-mass term.

First, there is one combination of the $H_p$ multiplets which remains
massless in the deconstruction phase. This is readily understood as we
have an odd number of chiral matter multiplets in the theory so at
least one chiral multiplet must remain massless. The zero-mode profile
is  
\beq 
\label{eq:zmplus}
H^{(0)} 
= 
\sqrt{\frac{\left({m}/{gv}\right)^2-1}
{\left({m}/{gv}\right)^{2N}-1}}
\sum_{p=1}^N \left (\frac{m}{gv} \right)^{p-1} H_p  
\eeq 
for the zero-mode of charge $+Q$ and  
\beq 
\label{eq:zmminus}
\tilde H^{(0)} 
= 
\sqrt{\frac{\left({gv}/{m}\right)^2-1}
{\left({gv}/{m}\right)^{2N}-1}}
 \sum_{p=1}^N \left (\frac{gv}{m} \right)^{p-1} 
\tilde H_p  
\eeq
for the zero-mode of charge $-Q$.
For $m < gv$ the $+Q$ ( $-Q$) zero-mode is localized near  the first
($N$-th) site, while for $m > gv$  it is localized near the $N$-th
(first) site .

The remaining combinations of $H$ and $\tilde H$  multiplets become
massive.  Their masses organize themselves  into a tower according to
the equation   
\beq
\label{eq:ms}
m_n^2 = g^2 v^2\left[(1 - \frac{m}{gv})^2 
+ 4\frac{m}{gv}\sin^2\left( \frac{n\pi}{2N} \right) \right], 
\eeq
while the decomposition of the $n$-th level massive mode is (only
formulae for the case of $+Q$ zero mode are given) 
\beq
\label{eq:4dwf}
H^{(n)} = \sqrt{\frac{2}{N}} \frac{gv}{m_n}  \sum_{p=1}^N \left [
 2 \sin \left(\frac{n\pi}{2N} \right) 
\cos \left( \frac{m\pi}{2N}(2p-1)\right)  
 + (m - gv) \sin  \left( \frac{n\pi}{N}\right)p \right]   H_p\,, 
\eeq
 \beq
\label{eq:4dwft}
\tilde H^{(n)} = \sqrt{\frac{2}{N}}  \sum_{p=1}^{N-1} 
\left [ \sin  \left( \frac{n\pi}{N}\right)p \right]  \tilde  H_p\,. 
\eeq

For $n \ll N$ and $|m-gv| \ll g v$  the mass formula (\ref{eq:ms})
becomes 
$m_n^2 \approx (m- gv)^2  + \frac{n^2\pi^2}{N^2}g^2 v^2$ which is
precisely the spectrum of a  5d hypermultiplet with the kink-mass $M
= (m- gv) $ for a compactification radius $1/R = \pi g v /N$. 
Thus the quantity $(m- gv)$ is related to the vev of the 5d singlet field 
$\Sigma$, which generates a kink-mass term in 5d, see eq. (\ref{e.kinkmass5d}).
 In this sense the vev $v$ of the links in deconstruction and the vev of the $\Sigma$ in 5d are related, $m -g\langle \Phi \rangle \to   \langle \Sigma \rangle$. 
The correspondence of the spectra holds for $ M \ll g v $, that is for a
kink-mass much smaller than the deconstruction scale (interpreted as
the cut-off $\Lambda$ of the 5d theory). For $M \sim gv $ the
massive spectra of the 5d and the deconstruction models differ
significantly. This could be expected, as deconstruction can
reproduce the features of the 5d theory only much below $\Lambda$.

One can also easily verify that the eigenmode decomposition is
analogous as in the 5d case. 
In this sense the correspondence between the physical space and the
group product space holds for deconstruction of orbifold
theories. Although the high-energy details are different in the two
theories, the 5d KK mode profiles can be mapped onto mode
decomposition in deconstruction and the precision of the mapping is
of the order of the `lattice spacing' $\dy = (gv)^{-1}$.

In 5d scenarios, apart from bulk hypermultiplets one often considers
chiral multiplets localized at the $\ZZ_2$ fixed points. It is rather
intuitive that the corresponding objects in deconstruction are chiral
multiplets charged under the first or the $N$-th group. Furthermore
such multiplets should not be coupled via the link-Higgs fields to
multiplets living at other sites; otherwise such multiplets would be
removed from the low-energy spectrum. 

The correspondence can be seen in a more formal way.     
Consider a multiplet $P$ charged under the $i$-th $U(1)$ group
only. This means it couples to the $i$-th gauge field, for example:    
\beq
S = \int \d^4x iq g P^\dagger \pa_\mu P A_{\mu,i}
\,.
\eeq
When the mode decomposition of the gauge field is inserted, this
coupling  becomes:  
\beq  
S = \int \d^4x\sum_{n=0}^{N-1} 
\sqrt {\frac{2}{N}} iq g P^\dagger \pa_\mu P A_\mu^{(n)} \eta_n 
\cos \frac{n(2i-1)\pi}{2N}            
\, ,
\eeq
where $\eta_n = 1/(\sqrt 2 ^{\delta{n0}})$.  
Now, recall that the analogous coupling of the chiral multiplet
localized at the brane at $y=y_i$ to the KK tower of the 5d gauge
field is: 
\beq 
\label{e.???}
 S = \int \d^4x\sum_{n=0}^\infty 
\frac{1}{\sqrt{\pi R}} iq g_5 P^\dagger \pa_\mu P A_\mu^{(n)} \eta_n
\cos \frac{n y_i}{R} 
\,.
\eeq   
We can see that in deconstruction a multiplet at the $i$-th site
couples analogously as  a brane field at the brane position 
$y_i = \frac{i-1/2}{gv}$. In particular, a multiplet at the first site
corresponds not exactly to the boundary multiplet at $y=0$ but to the
brane multiplet at $y=\frac{1}{2gv}$. Thus in deconstruction the
fixed point is resolved only up to the distance scale  
$\Delta y =\frac{1}{gv}$. One of the consequences of this is that it
is not possible to reproduce the effects of the $\de''$ terms present
in the 5d FI terms (\ref{e.fib}). Only exponential localization may
take place in deconstruction while in 5d models also a sharp
$\de$--like localization is possible.

We are now ready to discuss the question of stability of various
configurations of  `bulk' and `brane' matter multiplets which yield a
non-anomalous  spectrum of the zero modes, similarly as it was  done
in the 5d case. We start with the example of one replicated $(H_p,\t
H_p)$ multiplet with the $+Q$ charged zero mode (the $+Q$ charged
zero mode is obtained by removing $\tilde H_N$) and one brane
multiplet $P$  with charge $-Q$ living at the $N$-th site.    
In such configuration, the opposite in sign anomalies are localized
at both ends of the group product space (we assume that the mixed and boundary anomalies from the links are already cancelled by eq. (\ref{e.superwz})). Such globally vanishing anomaly can be cancelled via a WZ term but  in the case at hand we need a different WZ term than that of eq. (\ref{e.superwz}) since we do not want new mixed anomalies to be produced. It is straightforward to see that the  WZ term of the form 
\bea
\label{e.superwzd}
&
\cl_{SWZ} = -\frac{D}{24 \pi^2} Q^2 \int d^2\theta 
\log (\Phi_p/v) \left [
   W_{\alpha,p}W^{\alpha}_p +   W_{\alpha,p+1}W^{\alpha}_{p+1} 
+  W_{\alpha,p}W^{\alpha}_{p+1}  \right ]
& \nn & 
- \frac{D}{24 \pi^2}Q^2\int d^4\theta\left [ 
 (V_p D_\alpha V_{p+1} - V_{p+1} D_\alpha V_{p}) 
(W^{\alpha}_p + W^{\alpha}_{p+1})
\right ] +{\rm h.c.}&
\eea
reduces in the continuum limit to the CS term 
$\cl_{CS} = -\frac{D Q^2}{12 \pi^2}
\epsilon _{\alpha\beta\gamma\delta\epsilon} \left [
 A_\alpha \pa_\beta A_\gamma \pa_\delta A_\epsilon \right] $ 
and yields only boundary anomalies. 
Thus adding this WZ term to that of eq. (\ref{e.superwz}) and  choosing the constant $D$ appropriately we can cancel any boundary anomalies that globally sum up to zero. Since the continuum  theory contains the CS term, supersymmetry dictates that it also contains non-minimal  kinetic terms \cite{ARGRWA}.  
In consequence, in order to ensure the correct 5d Lorentz invariant continuum limit, the Kahler potential in deconstruction must also be supplemented with  non-minimal terms, see ref. \cite{dufapo}.

However cancellation of anomalies does not necessarily imply the stability of the configuration. If, locally, the charges do not sum to zero,  the FI terms will receive corrections at one loop. These corrections will respect the
relation $\xi_1 + \xi_N =0$ and supersymmetry will stay
unbroken. However, as the absolute value of the FI terms are shifted,
the link-Higgs vevs are shifted too and the fine-tuning between the
$m$ parameter  and $gv$  no longer holds. In the case at hand the net
charge in  the $N$-th group is negative, thus $\xi_N$ will decrease
and we obtain $m > gv$ at one loop. This result in an analogous effect
as that noticed in ref.\ \cite{GRNIOL}, namely, in spontaneous
localization of the zero-mode of  $H$ at the site where $P$
resides. Note however that the similar configuration, with $P$ living
at the first site  is perfectly stable, and so it does not correspond
to any 5d configuration.

The other examples of matter field configurations discussed in the 5d
case  also have their analogues in deconstruction. Two chiral
multiplets living at the different fixed points translate into one
chiral multiplet living at the first site and one chiral multiplet of
opposite charge living at the $N$-th site. Similarly as in the
previous case, FI terms generated at one loop will shift the
link-Higgs vev and may lead to spontaneous localization of any
replicated multiplets present. Another example is the case
with two replicated multiplets $(H_p, \tilde H_p)$ and $(G_p, \tilde
G_p)$. As discussed, non-anomalous zero-mode spectrum is obtained  by
removing $\tilde H_N$ and $G_1$. The net charge in the boundary group vanishes. For example, in the first group,  the $-Q$ charge of $\tilde G_1$ is balanced by  the $+Q$ charge of the link $\Phi_1$. Therefore, the configuration with two hypermultiplets of opposite zero-mode charges is stable, similarly as in 5d.  
The situation changes if we introduce the second pair of  replicated multiplets  $(H_p', \tilde H_p')$ and $(G_p', \tilde G_p')$ of opposite zero-mode charges.  This step leads to non-zero,
opposite in sign anomalies localized at the two endpoints of the
group product space. One can cancel these anomalies via WZ terms, but
again one loop corrections will destabilize the FI terms.  The
resulting shift of the link-Higgs vevs leads to localization of the
zero-mode  of $H$, $H'$ and $G$,$G'$ at the opposite endpoints. This situation is obviously different than in the 5d case. 
We would like to stress that these differences are not related to
the $\delta''$ term present in the 5d models. Such term is just absent
for arbitrary number of hypermultiplet pairs with opposite signs
(it is multiplied by a coefficient proportional to the sum
of all hypermultiplet charges).

More generally, one can formulate the condition for the stability of
the bulk fields in various  `bulk-brane' configurations in
deconstruction. Formulated at the level of the theory above the
deconstruction scale it says that charges of all chiral multiplets
have to cancell out locally in the group product space. We can also
easily rephrase this condition in terms of the effective theory below
the deconstruction scale. Denoting by $q_1$ ($q_N$) the charges of
chiral 
multiplets living on the first ($N$-th) site and by $q_+$, $q_-$ the
positive and negative (relatively to the sign of  $Q$) charges of the
zero modes of the  
replicated multiplets, the stability conditions are
\beq
\sum q_- + \sum q_1 +Q = 0
\,,\qquad\qquad
\sum q_+ + \sum q_N -Q = 0
\,.
\label{eq:stability}
\eeq     
{ As in the 5d case, they are more restrictive than merely the condition
for anomaly cancellation  $\sum q_- + \sum q_+ + \sum q_1 + \sum q_N = 0$. 
These two conditions ensure the cancellation of  quadratically divergent contributions to the FI terms in the boundary groups and are direct analogoues of the two conditions in eq. (\ref{eq:stability5}).  
We have checked by explicit calculation that, in the deconstruction framework,  finite and logarithmically divergent contributions to FI terms are subleading with respect to quadratically  divergent ones. This could be expected, as deconstruction provides  a regularization in which the brane thickness (set by the lattice spacing) is larger than the inverse cutoff scale of the theory, $\sigma \sim {1 \over gv} > \Lambda^{-1}$. 
Therefore an analogue of the  condition (\ref{eq:stability5s}) does not appear in  deconstruction. 



Note however the the two conditions (\ref{eq:stability}) are not exactly the same as the two first conditions in
(\ref{eq:stability5}). }
The bulk hypermultiplets in 5d enter in a fully symmetric way. This is not the case  in deconstruction where the positively charged
bulk fields are related to some of the brane fields while the
negatively charged bulk fields are related to the remaining brane
fields. This difference is due to the orientation of the group product
space which is determined by charges of the link fields. For some
aspects of the theory the end points of the aliphatic quiver diagram can 
be interpreted as the end points of the orbifold but for other aspects
they lose this interpretation. Then they are related rather to the two
possible signs of the U(1) charges.

\vspace{.3cm}
In summary, we have investigated deconstruction of 5--dimensional U(1) gauge
theories compactified on $S^1/\ZZ_2$ with various sets of bulk and
matter multiplets. We found that for most configurations of multiplets
the 5d theory has its deconstructed version. There are however also
some exceptions.

In both theories localized FI terms and localized anomalies 
(localized in the 5--th dimension or in the group product space,
respectively) are generated for general sets of bulk and brane matter
multiplets. If such would be anomalies sum up to zero in the effective
4d theory they can be cancelled by local counterterms without changing
the low energy spectrum. In 5d theories the Chern--Simons terms can be
used for that purpose. In deconstruction the corresponding mechanism
consists in adding Wess--Zumino terms.

{
In both types of models the necessary condition for unbroken
supersymmetry is the same as the condition for absence of anomalies: 
the sum of charges (in deconstruction, under the diagonal group) 
of all zero modes must be zero.
Another similarity between the 5d and deconstructed models is that in
both cases the localized FI terms lead to the localization of the bulk
fields. Typically the bulk fields of one sign of the charge of the
zero modes are localized around one of the branes while that with the
opposite sign are localized in the vicinity of the other brane.
In 5d models stability conditions  relate  charges of the brane fields at each of the branes with the sum of the zero mode charges of the bulk fields, as shown in  
eq.\ (\ref{eq:stability5}). If the $\delta''$ contributions to the FI terms are important then there is also the condition of eq.\ (\ref{eq:stability5s}) which puts a constraint on the sum of the bulk zero mode charges. 
 In deconstruction stability conditions  are given by 
eq.\ (\ref{eq:stability}) and relate charges of some of the brane
fields with  
positively charged bulk fields (and the remaining brane fields with
the negatively charged bulk fields).
 The UV completion provided by deconstruction corresponds to large brane thickness $\sigma  > \Lambda^{-1}$ and therefore an analogue of the stability condition  (\ref{eq:stability5s}) does not appear.

 The underlying reason for the difference between eq.\ (\ref{eq:stability5}) and eq.\ (\ref{eq:stability}) is the orientation of the
group product space in deconstruction which has no analogue in 5d
models.
In consequence, the 5d theory
with certain sets of multiplets, e.g. with two pairs of
hypermultiplets with zero-modes of opposite charges, does not have
its deconstructed 
analogue.
 Deconstruction can be considered a UV completion of
higher-dimensional theories and, as such, it should reproduce the
higher dimensional computations of UV insensitive observables. Our
analysis shows that stability conditions are sensitive to UV physics.     
}

\vspace{.3cm}
\section*{Acknowledgments}
We thank Emilian Dudas for pointing out the relevance of mixed anomalies. 

Work was supported in part by the European Community's Human Potential
Programme under contracts HPRN--CT--2000--00131 Quantum Spacetime,
HPRN--CT--2000--00148 Physics Across the Present Energy Frontier
and HPRN--CT--2000--00152 Supersymmetry and the Early Universe.
SP was partially supported by the Polish KBN grant 2 P03B 129 24
for years 2003--2004.
AF was partially supported by the Polish KBN grant 2 P03B 070 23
for years 2002--2004.

\end{document}